\documentclass[%
superscriptaddress,
twocolumn,
amsmath,amssymb,
aps,
pra
]{revtex4}

\usepackage{graphicx}
\usepackage{dcolumn}
\usepackage{bm,braket}
\usepackage{CJK}

\usepackage{color}
\begin{document}

\preprint{APS/123-QED}

\title{Quantum-classical correspondence in integrable systems}
\begin{CJK}{UTF8}{gbsn}
\author{Yiqiang Zhao(赵义强)}
\affiliation{
	International Center for Quantum Materials, School of Physics, Peking University, Beijing 100871, China
}
\author{Biao Wu(吴飙)}
\email{wubiao@pku.edu.cn}
\affiliation{
 International Center for Quantum Materials, School of Physics, Peking University, Beijing 100871, China
}
\affiliation{Collaborative Innovation Center of Quantum Matter, Beijing 100871, China}
\affiliation{Wilczek Quantum Center, School of Physics and Astronomy,\\
	Shanghai Jiao Tong University, Shanghai 200240, China}

\date{\today}

\begin{abstract}
We find that the quantum-classical correspondence in integrable systems is characterized by 
two time scales. One is the Ehrenfest time below which the system is classical; the other is the quantum revival time beyond
which the system is fully quantum. In between, the quantum system can be well approximated by classical 
ensemble distribution in phase space.  These results can be summarized in a diagram which we call Ehrenfest 
diagram. We derive an analytical expression for Ehrenfest time, which 
is proportional to $\hbar^{-1/2}$.  According to our formula,  the Ehrenfest time for the solar-earth 
system  is about $10^{26}$ times of the age of the solar system.
We also find an analytical expression for the quantum revival time, which is proportional to $\hbar^{-1}$. 
Both time scales involve  $\omega(I)$, the classical frequency as a function of classical action. 
Our results are numerically illustrated with two simple  integrable models. In addition, 
we show that similar results exist for  Bose gases,  where $1/N$ serves as an effective Planck constant. 
\end{abstract}

\maketitle
\end{CJK}

\section{\label{sec:level1}Introduction}
A quantum system is expected to become classical in the limit $\hbar\rightarrow 0$. 
However, how this exactly happens is highly non-trivial and has been studied intensively 
in the field of quantum chaos~\cite{Gutzwiller}. The issue of quantum-classical correspondence
was noticed as early as in 1927 by Ehrenfest. 
For a particle with mass $m$ moving in a potential $V(x)$, Ehrenfest  demonstrated that
the expectation values of the particle's position and momentum follow Newton-like equations~\cite{ehrenfest1927bemerkung}
    \begin{eqnarray}
            \frac{d}{dt}\langle \hat{x} \rangle &=& \frac{\langle \hat{p} \rangle}{m}   \\
            \frac{d}{dt}\langle \hat{p} \rangle &=& -\langle\frac{dV(\hat{x})}{d\hat{x}}\rangle
            \label{eq4}	    
    \end{eqnarray}
    where $\langle\cdot\rangle$ is the expectation value of the operator. These two equations are now known as 
    Ehrenfest theorem, which offers a hint on how quantum and classical dynamics may be related.     
    In particular,  when the wave function is narrow enough and/or the potential $V(x)$ varies gradually in space, 
    we approximately have 
    $\langle\frac{dV(\hat{x})}{d\hat{x}}\rangle\approx\frac{dV(\langle\hat{x}\rangle)}{d\langle\hat{x}\rangle}$. 
    This means that  the evolution of expectation values of position and momentum would follow exactly the Newton's 
    equation of motion. However,  an initially well-localized wave packet will spread, and the expectation 
    values of its position and momentum will eventually deviate from the classical dynamics when the width 
    of the wave packet is no longer small. Ehrenfest time $\tau_\hbar$ is  the time scale when 
    such a quantum-classical correspondence breaks down~\cite{berman1978condition,zaslavsky1981stochasticity,combescure1997semiclassical,hagedorn2000exponentially,berry1979evolution,silvestrov2002ehrenfest,tian2005,Fishman1984,Fishman1987,Lai1993}. 

 \begin{figure}
        	\centering
        	\includegraphics[width=0.75\linewidth]{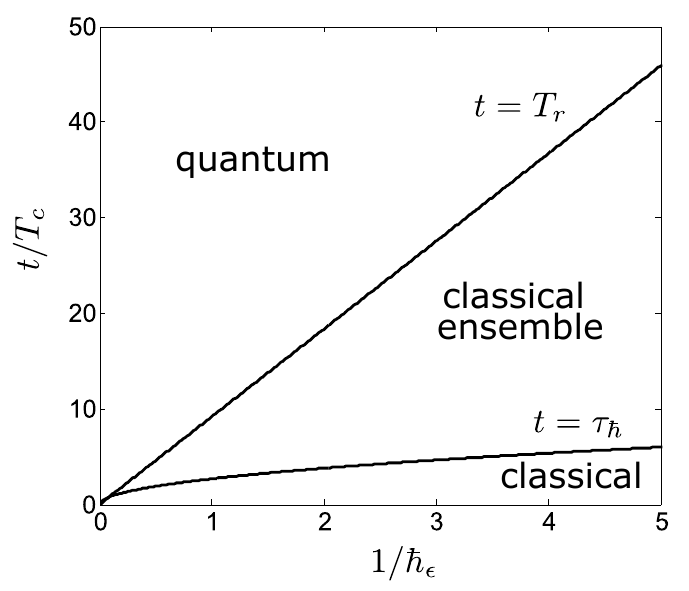}
        	\caption{Ehrenfest diagram for the quantum-classical correspondence in integrable systems. 
	Two time scales, Ehrenfest time and quantum revival time, 
	are plotted as functions $1/\hbar_\epsilon$.  These two times scales mark up three regions: quantum, classical ensemble, 
	and classical. The lines are plotted for the model in Eq.(\ref{oned}) with $(x_0=0, p_0=2)$ as the initial condition.}
        	\label{Ed}
        \end{figure}

In this work we study systematically the quantum-classical correspondence in integrable systems. 
We find that the quantum-classical correspondence is characterized by two time scales, Ehrenfest time 
$\tau_\hbar$ and quantum revival time $T_r$~\cite{robinett2004quantum,bakman2017collapse,veksler2015semiclassical}, as shown in Fig.\ref{Ed}. 
According to this figure, for a fixed Planck constant,  the wave packet dynamics is almost
classical when the evolution time is shorter than the Ehrenfest time $\tau_\hbar$;
when the evolution time is longer than $T_r$, quantum revival occurs and the wave packet
dynamics can no longer be approximated by semiclassical approaches. 
Between Ehrenfest time $\tau_\hbar$
and quantum revival time $T_r$,  the quantum dynamics can be well approximated by 
classical ensemble distribution in phase space.  
Furthermore, we are able to derive analytical expression for both Ehrenfest time $\tau_\hbar$
and quantum revival time $T_r$, both of which are intimately related to $\omega(I)$, the classical 
frequency as a function of classical action.  We find that $\tau_\hbar\propto\hbar^{-1/2}$ and 
$T_r\propto\hbar^{-1}$.


For many specific systems, we find that the Ehrenfest time has  a simple form $\tau_\hbar=cT_{c}(I/\hbar)^{1/2}$, where $T_{c}$ is the period of  a classical motion, $I$ is the corresponding action, and $c$ is a dimensionless constant of order one. 
Our results are applied to many concrete systems. Generally, for  systems which we usually regard as quantum systems, 
their Ehrenfest times are short; for systems which we usually consider as classical systems, their Ehrenfest times are long.
For example, for a hydrogen atom in the ground state, we have $\tau_\hbar=0.5T_c$; for the earth orbiting
around the sun, we have $\tau_\hbar=2.3\times 10^{36}T_c$ while the age of the solar system is only $0.5\times 10^{9}T_c$. 
Therefore, Ehrenfest time may be used as an indicator whether we should treat a given system as quantum or classical.


In the end we consider an integrable system of  Bose gas for which 
its effective Planck constant is  $1/N$~\cite{yaffe1982large}, where $N$ is the total number of the particle. 
When $N$ is small, the Bose gas is quantum and when $N$ is large it is well approximated by 
the mean-filed theory~\cite{han2016ehrenfest}. We also find two time scales, 
the Ehrenfest time scales with $N$ as $N^{1/2}$ and the quantum revival time scales linearly with $N$. 
As $N$ can be changed in an experiment, Bose gas offers a potential platform where the scalings of Ehrenfest time
and quantum revival time with the Planck constant may be verified  experimentally.

\section{\label{sec:level1}Ehrenfest time}
Before we present our general results, it is illuminating to look at concrete systems with numerical simulation. 
\subsection{numerical results}

We consider the following one dimensional  system   
        \begin{equation}
            H = \frac{p^2}{2m} + V(x)
            \label{oned}
        \end{equation}
where $m$ is the mass of the particle and $V(x)=m\omega_0^2x^2+m^2\omega_0^3 x^4/\hbar$. 
To numerically investigate how Ehrenfest time scales with the Planck constant, we set
the Planck constant in the Schr\"odinger equation as $\tilde{\hbar}=\hbar_\epsilon\hbar$, where 
the dimensionless constant $\hbar_\epsilon$ is varied.  In our numerical calculation, 
we use $\sqrt{\hbar/(m\omega_0)}$ as unit of length, $\sqrt{{\hbar m\omega_0}}$ as unit of momentum, 
$\hbar\omega_0$ as unit of energy, and  $1/\omega_0$ as unit of time. In this unit system, $V(x)=x^2+x^4$. 
        
    \begin{figure}[ht]
	    \centering
	    \includegraphics[width=7cm]{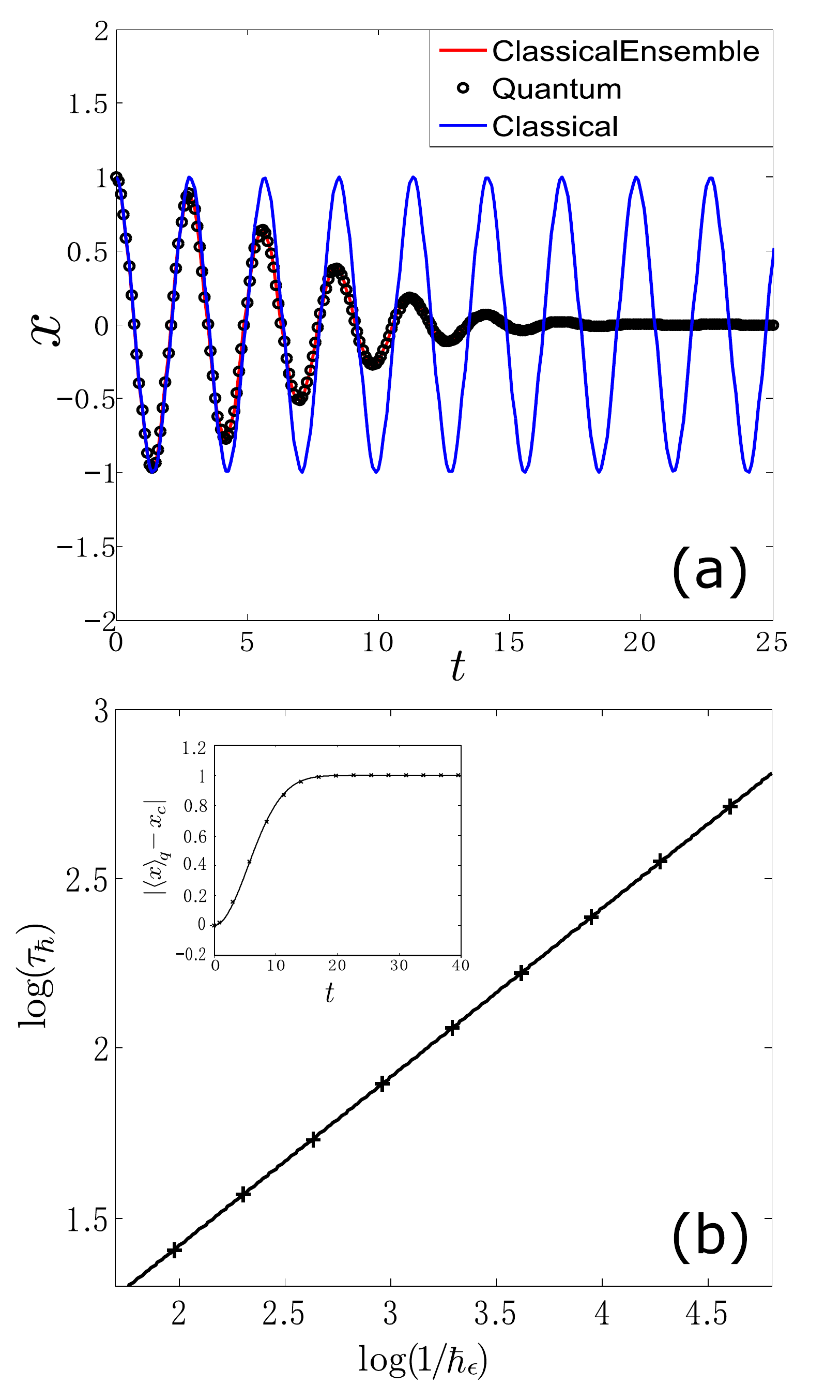}
    	\caption{(color online) (a) The time evolutions of a classical particle $x_c(t)$, its corresponding quantum expectation 
	value $\braket{x(t)}$, and the corresponding classical ensemble average $\bar{x}_c(t)$.  
	$x_0=1$, $p_0=0$, and $\hbar_\epsilon=0.03$. 
	(b) Relationship between between $\tau_\hbar$ and $1/\hbar_\epsilon$, which can be fit by function $y=0.5x+0.42$. 
	The inset is a typical fit curve of the evolution of the peaks of the difference between classical value and quantum expectation value. The unit of length is $\sqrt{\hbar/(m\omega_0)}$
	and the unit of time is $1/\omega_0$.}
	\label{qc}
   \end{figure}

We compare numerically the quantum and classical dynamics of this system.  For a given classical initial condition ${x_0,p_0}$, 
we construct  the following  Gaussian wave packet as the initial state for the quantum dynamics,
\begin{equation}
\label{gaussian}
\psi(x) = \frac{1}{(2\pi\sigma_x^2)^{1/4}}\exp\left\{-\frac{(x-x_0)^2}{4\sigma_x^2}+\frac{ip_0(x-x_0)}{\hbar_\epsilon}\right\}\,,
\end{equation}
where $\sigma_x=\sqrt{\hbar_\epsilon/2}$.  The quantum expectation value $\langle x(t)\rangle$ 
and  the classical trajectory $x_c(t)$ are compared in Fig. \ref{qc}(a).  As expected, they match each other for an initial short period of time and then start to deviate.  We find that the difference $|\langle x(t)\rangle-x_c(t)|$ oscillates and its peaks  can be approximated by function $y=a(1-e^{-bt^2})$, as shown in the inset of Fig. \ref{qc}(b). The Ehrenfest time is extracted from these numerical results as $\tau_\hbar=\sqrt{1/b}$. 
When $\hbar_\epsilon$ is varied, $\tau_\hbar$ varies. Their relation is shown in Fig. \ref{qc}(b), which clearly shows  $\tau_\hbar\propto \hbar^{-1/2}$.

\begin{figure}[!t]
	\centering
	\includegraphics[width=8cm]{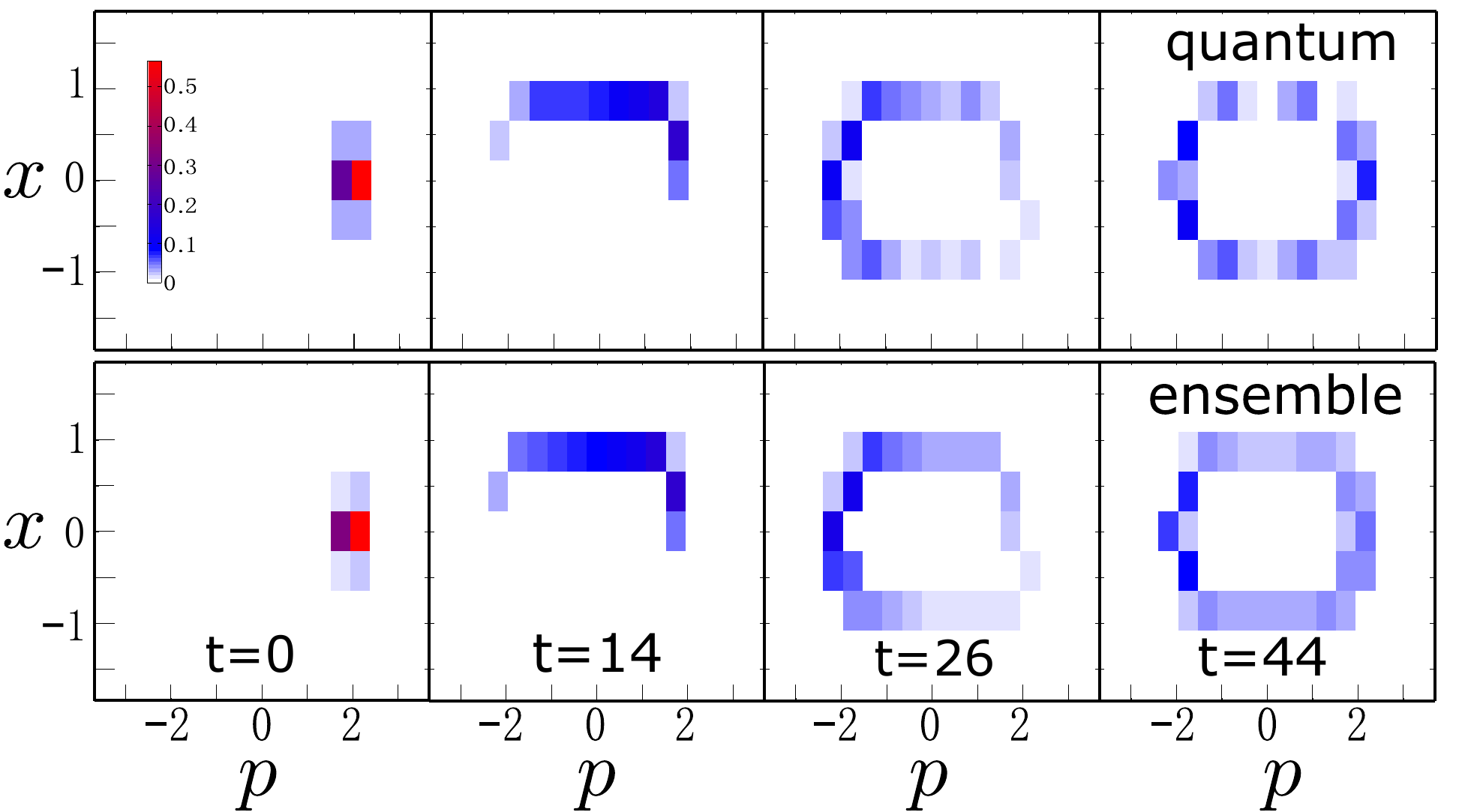}
	\caption{
		(color online) Quantum dynamics (upper row) and classical ensemble dynamics (lower row) in phase space. 
		The classical ensemble distribution 
		is coarse-grained to Planck cells. 	For this case, $\tau_\hbar\approx 26$. $(x_0=0,p_0=2)$ and $\hbar_\epsilon=0.03$.
		The unit of length is $\sqrt{\hbar/(m\omega_0)}$, 
		the unit of time $1/\omega_0$,  and the unit of momentum $\sqrt{{\hbar m\omega_0}}$. }
	\label{phase1}
\end{figure}

In addition,  we follow  Ref.~\cite{ballentine1994inadequacy} and compare the quantum dynamics to
its corresponding classical ensemble evolution. We use the Wigner function of the Gaussian wave packet in 
Eq.(\ref{gaussian}) as the initial distribution for a classical ensemble
      \begin{equation}
      \label{ensemble}
      \rho_c(x,p) = \frac{1}{\pi\hbar_\epsilon}\exp\left\{-\frac{(x-x_0)^2}{2\sigma_x^2}-\frac{(p-p_0)^2}
      {2\sigma_p^2}\right\}\,.
      \end{equation}
        where $\sigma_x=\sigma_p=\sqrt{\hbar_\epsilon/2}$. We use $\bar{x}_c$ as the classical ensemble average 
of $x$.  The agreement between the quantum expectation value $\langle x(t)\rangle$  and $\bar{x}_c(t)$ is almost perfect
for a short period of time  as shown in Fig. \ref{qc}(a). 
Such an excellent agreement goes beyond just the averaged value and exists even in phase space.  
To plot the quantum dynamics in phase space, 
we use the method in Refs.\cite{han2015entropy,Fang2017} to project wave function unitarily onto 
quantum phase space.  Roughly,  the classical phase space is divided into Planck cells
and each Planck cell is assigned a Wannier function; these Wannier functions form a complete orthonormal basis
which is used for the unitary projection. The results are plotted  in Fig. \ref{phase1}, where we see that 
the agreement is excellent within Ehrenfest time and it begins to break only after $t=26$. 

	To illustrate that our results hold for higher dimensions, we consider an integrable model of two degrees of freedom.
	It is a model constructed from three-site Toda lattice~\cite{toda1967vibration} with the following Hamiltonian
	\begin{equation}
	    H=\frac{p_1^2}{m} + \frac{p_2^2}{m} + \frac{p_1p_2}{m} + \mu \big[e^{-x_1/a}+e^{-(x_2-x_1)/a}+ e^{x_2/a}\big]\,.
	\end{equation}
	Similarly, we set
	the Planck constant in the Schr\"odinger equation as $\tilde{\hbar}=\hbar_\epsilon\hbar$, where 
	the dimensionless constant $\hbar_\epsilon$ can be varied.  In our numerical calculation, 
	we use $a$ as unit of length, $\sqrt{m\mu}$ as unit of momentum, 
	$\mu$ as unit of energy, and  $\hbar/\mu$ as unit of time. In this unit system, $H=p_1^2+p_2^2+p_1p_2 +e^{-x_1}+e^{-(x_2-x_1)}+e^{x_2}$. 
	Two independent conserved quantities are $H$ and $F=-(p_1^2p_2+p_2^2p_1)-p_2e^{-x_1}+(p_1+p_2)e^{-(x_2-x_1)}-p_1e^{x_2}$.
	
	\begin{figure}[ht]
		\centering
		\includegraphics[width=7cm]{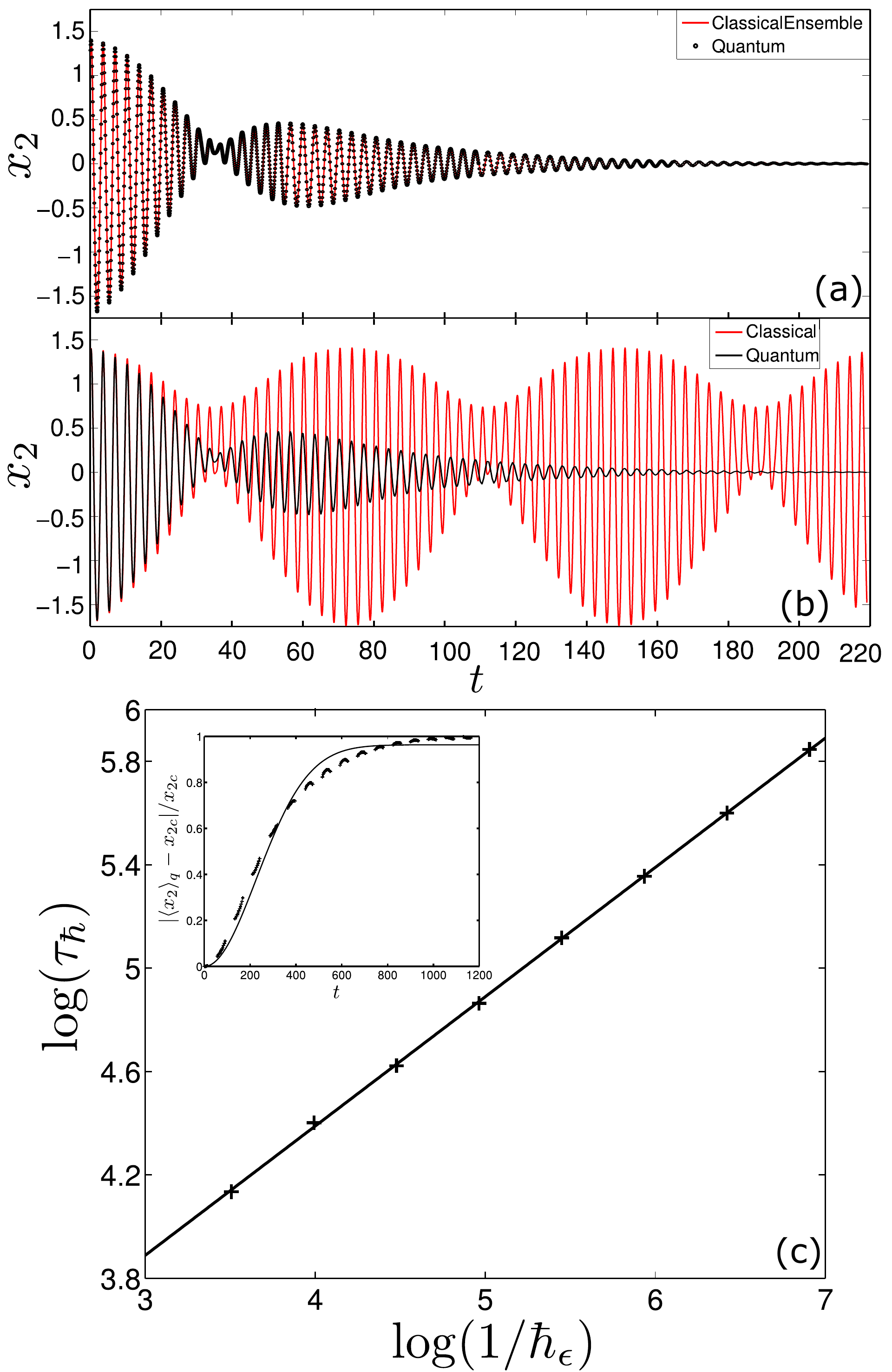}
		\caption{
				(color online) Numerical results for the integrable model of two degrees of freedom. (a)(b) The time evolutions of a classical particle position  $x_{2c}(t)$, its corresponding quantum expectation 
			value $\braket{x_2(t)}$, and the corresponding classical ensemble average $\bar{x}_{2c}(t)$.  The initial condition is
			$x_1=0.7, x_2=1.2, p_1=0.4, p_2=0.6$. $\hbar_\epsilon=0.03$. 
			(c) Relationship between between $\tau_\hbar$ and $1/\hbar_\epsilon$, which can be fit by function $y=0.50x+2.39$. The inset is a typical fit curve for the evolution of relative difference between classical value and quantum expectation value $|x_{2c}(t)-\langle x_2(t)\rangle|/x_{2c}(t)$. The unit of length is $a$ and the unit of time is $\hbar/\mu$.}
		\label{2dqc}
	\end{figure}
     
     The computation procedure is similar to the one dimensional case. To determine Ehrenfest time numerically, we use relative difference $|x_{2c}(t)-\langle x_2(t)\rangle|/x_{2c}(t)$ as a criterion. To avoid zero points of $x_{2c}(t)$, we choose time points when $x_{2c}(t)$ is large. As showed in Fig.~\ref{2dqc}, the Ehrenfest time in the two dimensional system also scales with $\hbar$ as 
     $\tau_\hbar\propto \hbar^{-1/2}$.


 \subsection{General analysis}
The  numerical results above also indicate that a single-particle classical trajectory
deviates from its corresponding classical ensemble dynamics (see Fig. \ref{qc}(a) and Fig. \ref{2dqc}(a)(b)), which was already noticed
in Ref.\cite{ballentine1994inadequacy}. This fact, together with the perfect agreement between quantum dynamics and 
classical ensemble dynamics within Ehrenfest time, implies that Ehrenfest time $\tau_\hbar$ 
is solely caused by the width of a quantum wave packet that has a lower limit set by the uncertainty relation. 
We exploit it to derive an analytical expression for Ehrenfest time. 

We consider a classical ensemble distribution that satisfies the uncertainty relation, such as the one in Eq.(\ref{ensemble}). 
The evolution  of this classical ensemble is governed by Liouville equation, which is  totally classical and irrelevant of $\hbar$. 
The only factor related to $\hbar$ is the fluctuations of position and momentum in this ensemble  distribution 
which are limited by the uncertainty principle.

 \begin{figure}[h]
       	\centering
       	\includegraphics[width=6cm]{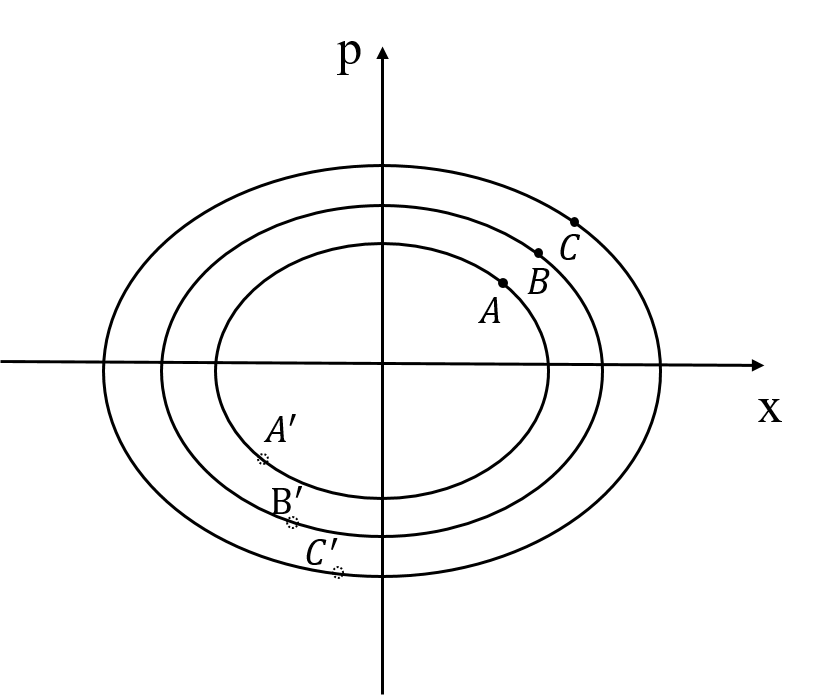}
       	\caption{A typical phase space for one dimensional integrable system. Closed curves are energy contours. In general, the oscillation frequencies on different curves are different. So, the three points A, B, and C , initially close to each other, 
	will disperse over time due to different fequencies. }
       	\label{abc}
        \end{figure}

 We choose three points A, B, and C in the phase space such that they initially differ from each other by 
 $\delta p$  in  momentum  and  $\delta x$ in position (see Fig.\ref{abc}).  In particular, B is the averaged point
 of A and C.  As long as the system is not a harmonic oscillator, these three points have different angular velocities. 
 As time goes by,  the average of A and C  will differ significantly from B and 
 the correspondence between classical ensemble and classical single particle will break down. 
 When we choose $\delta x\cdot\delta p\sim \hbar$, such a breakdown time is just Ehrenfest time $\tau_\hbar$. 
 
 We define $\tau_\hbar$ as the time when the angular difference of A and C is $2\pi$. We thus have
        
 \begin{eqnarray}       
     \tau_\hbar &= &\frac{2\pi}{|\omega_A-\omega_C|}\\
      &\approx&\frac{2\pi}{|\omega^\prime(I)|\cdot (|\partial I/\partial x|\cdot\delta x+|\partial I/\partial p|\cdot\delta p)}
     \label{tau}
\end{eqnarray}

where $\omega$ is the angular velocity and $I$ is the action. Note that all these quantities $\partial I/\partial p$, $\partial I/\partial x$ and $\omega^\prime(I)$ are classical  and independent of $\hbar$.  The Planck constant comes in only through
the uncertainty relation that requires that $\delta x\sim \delta p\propto\hbar^{1/2}$. So, we have
        \begin{equation}
            \tau_\hbar\propto \hbar^{-1/2}\,.
            \label{Eh}
        \end{equation}

There is no need to worry about the possibility $\omega^\prime(I)=0$ in Eq.(\ref{tau}) as it 
is the result of truncation of  the Taylor expansion of $|\omega_A-\omega_C|$  to the first order.  If $\omega^\prime(I)=0$, 
one just needs to expand it further to the second order. In this case, we would have $\tau_\hbar\propto \hbar^{-1}$. 
One could continue this expansion until some order becomes non-zero. 
If all orders of derivative of $\omega(I)$ vanish, the system must be 
a harmonic oscillator for which $\tau_{\hbar}$ is indeed infinite.

	For n-dimensional integrable system, there exist $n$ pairs of independent action-angle variables and thus $n$ angular velocities, each of which gives a Ehrenfest time according to Eq. (\ref{tau}), 
	 \begin{equation}
    \tau_{\hbar i}=\frac{2\pi}{|\partial\omega_i/\partial I_j|\cdot (|\partial I_j/\partial x_k|\cdot\delta x_k+|\partial I_j/\partial p_k|\cdot\delta p_k)}
    \end{equation}
    where $i,j,k = 1, 2, ...,n$, and repeated indices imply summation. In phase space, the spread of a wave packet in any direction will cause the break down of the quantum-classical correspondence, so the shortest of them will be Ehrenfest time for the $n$-dimensional integrable system, that is, 
    \begin{equation}
    \tau_\hbar = \min\{\tau_{\hbar i}\}\,,\quad\quad (i = 1, 2, \cdots,n)
    \label{ndtau}
    \end{equation}

For chaotic systems, it is well accepted that Ehrenfest time $\tau_\hbar=\frac{c}{\gamma}\ln{\frac{A}{\hbar}}$~\cite{berman1978condition,zaslavsky1981stochasticity,combescure1997semiclassical,hagedorn2000exponentially,berry1979evolution,silvestrov2002ehrenfest,tian2005}, where $\gamma$ is the Lyapunov index of the chaotic system, $A$ is a typical action, and $c$ is a dimensionless constant of order one. 
However, there is some confusion over Ehrenfest time in integrable systems. 
Although  it is generally believed that  for integrable systems Ehrenfest time scales 
with the Planck constant as  $\tau_\hbar\propto\hbar^{-\alpha}$~\cite{berman2008crossover}, it is not clear
in literature what $\alpha$ is. 
It was indicated
in Ref.~\cite{Fishman1987} that $\alpha=1$ but no detailed explanation was given. 
However, it is shown in some specific cases  that $\alpha = 1/2$ \cite{berman2008crossover,berman1981method}.  
Berry and Balazs studied a similar time scale with Wigner function and found that $\alpha=2/3$~\cite{berry1979evolution}. 
Our work clarifies this issue and shows analytically $\alpha=1/2$. 
Note that Ehrenfest time is intrinsic to the system and is independent of initial conditions. It can be understood 
as the time scale that a classical ensemble distribution in phase space develops structures finer than Planck cell.

\subsection{Examples}
\label{sec:ex}
 We now apply the above result to a couple of examples to get a sense how big or small the Ehrenfest time can become in
 typical macroscopic and microscopic situations.  The first example is a particle of mass $m$ in a one dimensional box of length  $a$. Through some simple calculations we have
        \begin{eqnarray}
            \tau_\hbar&=&T_c\sqrt{\frac{2I}{\hbar}}
        \end{eqnarray}
        where $I = pa/\pi$ is the action and $T_c = 2 a m/p$ is the classical period with $p$ being the momentum of
        the particle. Here we consider two typical scenarios, one macroscopic and one microscopic.  Imagine that a macroscopic ball moves in a box with $m=1$g, $a=1$m, $v=1$m/s. The Ehrenfest time for this system is
 then $\tau_\hbar = 2.4\times 10^{15} T_c$. Naturally, classical mechanics is enough to describe such a system. 
 For the microscopic scenario, we consider a ultracold $^{87}$Rb atom moving in a optical well~\cite{coldatom}, where 
 $m = 1.5\times10^{-25}$kg, $v=10^{-3}$m/s (estimated under condition $T=10^{-8}$K), and $a=10^{-7}$m (roughly the wavelength of light). The Ehrenfest time for this case is $\tau_\hbar = 0.8T_c$. So ultracold  atoms must be described by quantum mechanics. This example shows that Ehrenfest time is a good indicator whether a system should be regarded 
 as quantum or classical.
        
        The second example is a system with the inverse square law of force, whose Hamiltonian is
        \begin{equation}
            H = \frac{p_r^2}{2m}+\frac{L^2}{2mr^2}-\frac{k}{r}
        \end{equation}
        where $m$ is the mass of the object, $r$ is the distance to the center, $p_r$ is the radial momentum, and $L$ is the angular momentum. With canonical transformation, we have
        \begin{equation}
            \begin{aligned}
                H = -\frac{mk^2}{2}\frac{1}{(I+L)^2}
            \end{aligned}  
        \end{equation}
        where $I$ is the  action variable of the system other than $L$. To simplify the calculations, 
        we choose a special initial condition $r = \frac{L^2}{mk}$, and the variances of the wave packet are
        $\delta r = \sqrt{\hbar/(2m\omega)}$,  $\delta p_r = \sqrt{m\omega\hbar/2}$, $
        \delta \theta = \frac{1}{r}\sqrt{\hbar/(2m\omega)}$, and $\delta L = r\sqrt{m\omega\hbar/2}$.         
        With some simple calculations we have
        \begin{eqnarray}
        \tau_\hbar&=&\frac{\sqrt{2}}{3}\frac{T_c}{\sqrt{\frac{\hbar[(I+L)^2-L^2]}{L^2(I+L)}} + \frac{L}{I+L}\sqrt{\frac{\hbar}{L}}}
        \end{eqnarray}
For the sun-earth system, as the motion is approximate circular motion,   we have $I \approx 0$ and 
$L=2.7\times10^{39}J\cdot s$. So, we have $\tau_\hbar=2.3\times10^{36}$ years 
while the age of the solar system is just $5\times 10^9$ years. For a hydrogen atom in its ground state, as $L=\hbar$ 
we have $\tau_\hbar=0.5T_c$.  This is clearly consistent with our daily experience that
we do not need to worry about the quantum effects in the orbits of the solar planets 
while we have to describe  hydrogen atom with quantum mechanics.

\section{Quantum revival time}     
Ehrenfest time gives us the time scale when the quantum dynamics of a single particle deviates from its 
classical trajectory. However, as shown in Fig.  \ref{qc}\&\ref{2dqc}, if one compares the dynamics of a quantum wave packet 
to an ensemble of classical orbits, the quantum-classical correspondence can last much longer. This phenomenon
of course has been noticed a long time ago~\cite{ballentine1994inadequacy}. In this section, 
we investigate how long the quantum-classical correspondence can last in this sense. We find that for integrable systems
such a time scale  is set by quantum revival~\cite{robinett2004quantum,bakman2017collapse,veksler2015semiclassical}
and scales with the Planck constant as $\hbar^{-1}$.

        \begin{figure}[!t]
        	\centering
        	\includegraphics[width=0.9 \linewidth]{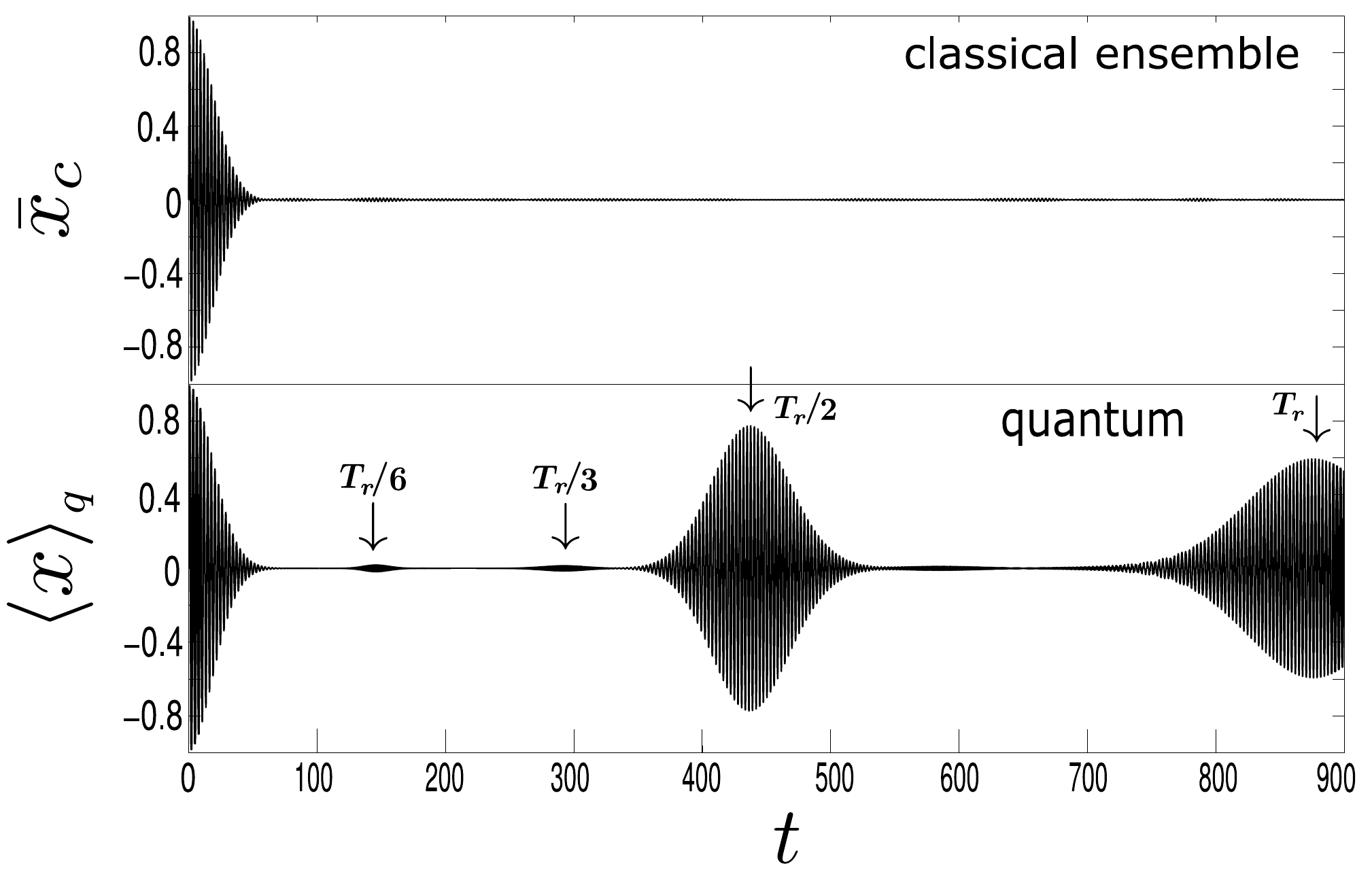}
        	\caption{(Upper) Time evolution of the classical 
	ensemble average $\bar{x}_c(t)$; (Lower) time evolution of the quantum expectation value $\braket{x(t)}$. 
	$T_r$ is the quantum revival time.  $(x_0=0,p_0=2)$ and $\hbar_\epsilon=0.03$. The unit of length is $\sqrt{\hbar/(m\omega_0)}$ and the unit of time $1/\omega_0$. }
        	\label{ensemble}
        \end{figure} 
          \begin{figure}[ht]
        	\centering
        	\includegraphics[width=8cm]{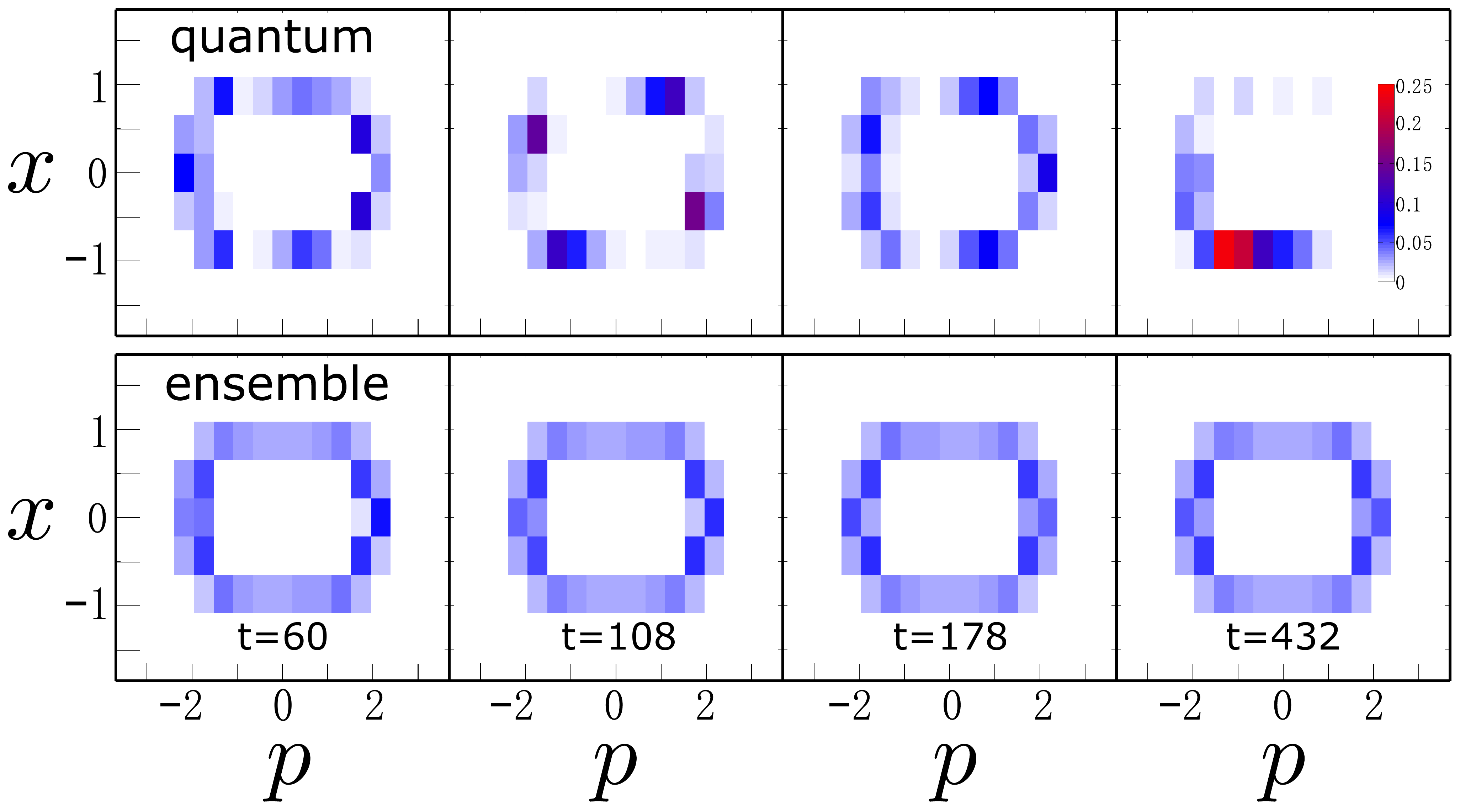}
        	\caption{(color online) Quantum dynamics (upper row) and classical ensemble dynamics (lower row) in phase space. 
	The classical ensemble distribution 
	is coarse-grained to Planck cells. 	For this case, $T_r\approx 864$. $(x_0=0,p_0=2)$ and $\hbar_\epsilon=0.03$.
	The unit of length is $\sqrt{\hbar/(m\omega_0)}$, 
	 the unit of time $1/\omega_0$,  and the unit of momentum $\sqrt{{\hbar m\omega_0}}$. }
        	\label{phase2}
    \end{figure}

\subsection{Numerical results}
        
We further study numerically the quantum dynamic and its corresponding classical ensemble dynamics for much longer times. 
They are compared in term of  the averaged position (see Fig. \ref{ensemble}) and also in phase space (see Fig. \ref{phase2}).  
If one is only interested in the dynamics of the wave packet center, the quantum and classical ensemble results
match each other very well for a very long time, up to $t>300$ according to Fig. \ref{ensemble}.  After that, 
around $t\approx 430$, while the classical average $\bar{x}_c(t)$ remains around zero, the quantum 
expectation $\braket{x(t)}$ almost fully recovers its original value, which is known as quantum revival. 
This quantum revival occurs again when the evolution time is doubled. 

However, if one is interested in more dynamical details, the time scale of agreement is shortened by a few fractions. 
According to Fig. \ref{phase2},  after $t=\tau_\hbar\approx 26$, both are no longer localized. However, 
the quantum distribution always has more structures while the classical ensemble distribution is rather uniformly distributed
within the energy shell. In particular, at certain times, one observes that   the  quantum distribution will cluster
around a few centers, a phenomenon known as  fractional quantum revival~\cite{robinett2004quantum}. 
At $t=432$, which is half of the quantum revival time, we see that   the quantum  distribution becomes localized again.

 \subsection{Analytical results}    
The numerical results above show that  quantum 
dynamics and its corresponding classical ensemble dynamics begin to deviate from each other
significantly when  quantum revival occurs . In this subsection, 
we derive an analytic formula for quantum revival time. We follow the method in Ref.~\cite{robinett2004quantum} 
but with a significant modification by introducing action variables.  For a general one dimensional integral system,  
its classical Hamiltonian can always be written as $H(I)$, where $I$ is the action of the system. As a result, its classical 
energy is also a function of the action $E(I)$ and so is the classical frequency $\omega(I)=\partial E(I)/\partial I$~\cite{Arnold1978Book}. We expand the quantum wave packet in terms of the system's energy eigenstates 
and its dynamics is then given by 
\begin{equation}
\psi(t)=\sum_n c_n e^{-iE_nt/\hbar}\phi_n(x)\,,
\end{equation}
where $\phi_n(x)$ is the $n$th eigenstate and $E_n$ is its corresponding energy eigenvalue. The coefficients $c_n$'s 
are determined by the initial condition. We assume that $|c_m|^2$ is the largest and expand
the eigenvalue around $E_m$ as  follows
        \begin{equation}
            E_n = E_m + \omega(I_m)(I_n - I_m) + \frac{\omega^\prime(I_m)}{2}(I_n - I_m)^2 +...
            \label{expansion}  
        \end{equation}
        where $I_n$ is the  action corresponding to $E_n$ via  $E(I)$. 
        According to the Bohr-Sommerfeld quantization rule~\cite{Messiah}, we have 
        \begin{equation}
             I_n - I_m = (n-m)\hbar\,.
             \label{bohr}
        \end{equation}
        So, the quantum  phases can be written as
        \begin{eqnarray}      
                & & \exp\left[-i(E_n-E_m)t/\hbar\right] \nonumber\\
                  &=&\exp\left[-2\pi i(n-m)\frac{t}{T_c}
                  -2\pi i(n-m)^2\frac{t}{T_r} + ...\right]\,,
                  \label{phase}
        \end{eqnarray}
        where $T_c = 2\pi/\omega(I_m) $ and 
        \begin{equation}
    	       T_r = \frac{4\pi}{\omega^\prime(I_m)\hbar}\,.
    	       \label{revival}
         \end{equation}
As $T_r$ contains $\hbar$ in its denominator, it is clear that $T_r\gg T_c$. With this in mind, we can envision 
from Eq.(\ref{phase}) how the quantum wave packet will evolve in time.  For an initial short interval of time, 
the wave packet will oscillate with period $T_c$ but with a decaying amplitude due to the second-order and other higher order 
terms.  When the evolution time approaches $T_r$, the second-order terms become multiples of $2\pi$
and, as a result, the wave packet recovers most of its original shape. How much it can recover depends on 
the third and higher order terms and other factors. Before $T_r$, there can be fractional quantum revivals that
occur at $t=pT_r/q$($p$, $q$ are positive integers); they are characterized by  a superposition of several localized wave packets~\cite{robinett2004quantum}. This is exactly what we have observed in Fig. \ref{phase2}.

From the above discussion, we find that  the quantum revival time $T_r$  scales with the Planck constant as $T_r\propto\frac{1}{\hbar}$. For the two examples mentioned in Sec. \ref{sec:ex}, according to Eq. (\ref{revival}), we have
    \begin{equation}
         T_{r1}=T_c\frac{2I}{\hbar}\,,
    \end{equation}
    and 
    \begin{equation}
         T_{r2}=\frac{2}{3}T_c\frac{I+L}{\hbar}\,,
    \end{equation}
  respectively.

We note that the Bohr-Sommerfeld quantization rule is only an approximation; Eq.(\ref{bohr})  should be 
corrected to $I_n - I_m = (n-m)\hbar + \delta_e$. For the above analysis to be correct, the condition $\omega(I_m)\delta_e\ll\frac{\omega^\prime(I_m)}{2}(n-m)^2\hbar^2$ should be satisfied.  $\delta_e$ also affects how much
the quantum wave packet can recover its original shape at $T_r$.

       
 \section{Bose gases}
 \begin{figure*}
        	\centering
        	\includegraphics[width=14cm]{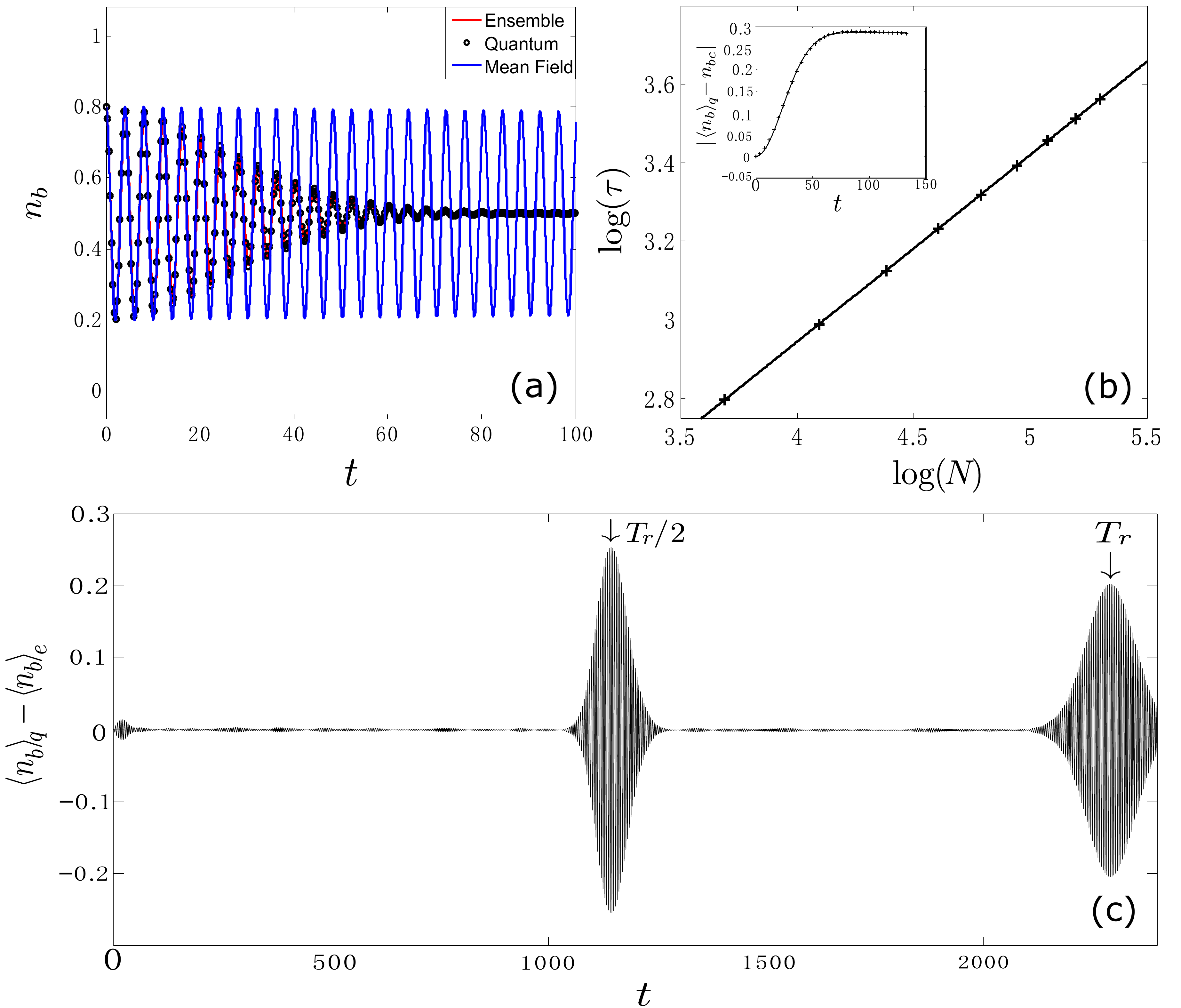}
        	\caption{(color online) (a) The time evolution of the averaged probability of the system in state $b$ according to 
	three different dynamics: quantum, mean-field,  and mean field ensemble.  (b) Ehrenfest time as a function of 
	number of particles $N$, which can be fit by $y=0.48x+1$.  
	(c) The evolution of difference between quantum expectation value and mean field ensemble average 
	of occupation probability at  state $b$.  $N=200$, $c/\nu=2$. The unit of time is $\hbar/\nu$}.
        	\label{bosegas}
        \end{figure*}

 It is well known that the relationship between quantum and mean field descriptions of Bose gases is  essentially 
 quantum-classical correspondence~\cite{yaffe1982large,han2016ehrenfest,veksler2015semiclassical} with  $1/N$($N$ is the total number of bosons) 
 serves as effective Planck constant. Our results above can be straightforwardly applied to any system of Bose gas which
 is integrable as it was done for chaotic Bose system in Ref. ~\cite{han2016ehrenfest}.  
 We illustrate this with a two-site Bose-Hubbard model as an example, whose  Hamiltonian is
        \begin{equation}
            \hat{H}=-\frac{\nu}{2}(\hat{a}^\dagger\hat{b}+\hat{a}\hat{b}^\dagger)+
            \frac{c}{2N}(\hat{a}^\dagger\hat{a}^\dagger\hat{a}\hat{a}
            +\hat{b}^\dagger\hat{b}^\dagger\hat{b}\hat{b})
        \end{equation}
        where with $\hat{a}^\dagger(\hat{a})$ and $\hat{b}^\dagger(\hat{b})$ the creation (annihilation) operators in well a and b, $c$ is the strength of interaction and $\nu$ is the tunneling parameter. In our numerical calculation, we use $\nu$ as unit of energy, $\hbar/\nu$ as unit of time. When the particle number $N$ is large,  
        this system can be well approximated by the following mean field model
        \begin{equation}
            H_{\rm mf}=-\frac{\nu}{2}(a^*b+ab^*)+\frac{c}{2}(|a|^4+|b|^4)\,.
        \end{equation}
Owing to the particle number conservation, $|a|^2+|b|^2=1$, and the overall phase is trivial, we can introduce a pair of 
conjugate variables $s$ and $\theta$, where $s=|b|^2$, $\theta=\theta_b-\theta_a$ with $\theta _b$ and $\theta_a$  being
the phases of complex numbers $a$ and $b$. The mean field model is clearly a classical one dimensional integrable system.
        
 In the above discussion of quantum-classical correspondence of a single particle, a point in the classical phase space
 corresponds to a  Gaussian wave packet of minimal spread. For this Bose system, a mean field state $a=\alpha$, $b=\beta$
 corresponds to a  quantum  coherent state$|\alpha, \beta\rangle$
        \begin{equation}
            |\alpha,\beta\rangle=\frac{1}{\sqrt{N!}}(\alpha a^\dagger+\beta b^\dagger)^N|0\rangle\,.
        \end{equation}
where $|0\rangle$ is the vacuum state.

However, we need some effort to construct  the corresponding mean field ensemble  distribution $\rho(s, \theta)$.
We expand the coherent state $|\alpha,\beta\rangle$ with Fock states $|n, N-n\rangle$, where $n$ is the particle 
number at site $a$, 
        \begin{equation}
            |\alpha,\beta\rangle = \sum_s \varphi_N (s) |N-Ns, Ns\rangle \,,
        \end{equation}
        where 
        \begin{equation}
            \varphi_N (s) = \sqrt{\frac{N!}{Ns!(N-Ns)!}}\alpha^{N-Ns}\beta^{Ns}\,.
        \end{equation}
        and $s$ ranges over  $0/N,1/N,...,N/N$. $ |\varphi_N (s)|^2$ can be regarded as
        a distribution of $s$. For this distribution, the average of $s$ is $ \bar{s} = |\beta|^2$ and its
        variance is 
        \begin{equation}
              \Delta s = \frac{|\beta|\sqrt{(1-|\beta|^2)}}{\sqrt{N}}\,.
        \end{equation}
        As $\theta$ is the conjugate of $s$,  its distribution can be obtained with a Fourier transform 
        \begin{equation}
                           \phi_N(\theta) = \frac{1}{\sqrt{N+1}}\sum_s\varphi_N(s)e^{-iNs\theta}  
        \end{equation}
        where $\theta$ takes the following discrete values: $2\pi\frac{1}{N+1},2\pi\frac{2}{N+1},...,2\pi\frac{N+1}{N+1}$. Numerical results show that
        \begin{equation}
            \bar{\theta} \approx \theta_\beta-\theta_\alpha  \,,\quad\quad  
             \Delta\theta \approx \frac{1}{2\sqrt{N}|\beta|\sqrt{(1-|\beta|^2)}}\,.
        \end{equation}
        
 So,   $\Delta\theta$ and $\Delta s$ satisfy the uncertainty relation:$\Delta\theta\Delta s\approx\frac{1}{2N}$. 
 At the large $N$ limit, $N\rightarrow+\infty$, both $|\varphi_N (s)|^2 $ and $|\phi_N(\theta)|^2$ will approach 
 Gaussian distribution. If we denote these two Gaussian distributions as $g_1(s)$ and $g_2(\theta)$, respectively, 
the mean-field ensemble distribution can be constructed as $\rho(s,\theta)=g_1(s)g_2(\theta)$. The 
three different dynamics, mean-field, mean-field ensemble, and quantum, are compared in Fig.~\ref{bosegas}. 
We find a very similar pattern as we found in Sections III and IV.  

For quantum revival, we would need
the Bohr-Sommerfeld quantization rule. How to implement this rule in the mean field theory of a Bose gas is
discussed in Ref.\cite{WuLiu,Luo}.

        
        In conclusion, the breakdown of correspondence between quantum and mean field descriptions occurs at time $\tau_\hbar\propto N^{1/2}$, and the breakdown of correspondence between quantum and mean field ensemble occurs at time $T_r \propto N$. 
        The Planck constant $\hbar$ can not be changed experimentally, but total number of bosons $N$ can. Therefore,  the  Bose gases can be used to experimentally verify the results in this paper.

\section{\label{sec:conc}Discussion and Conclusion}
    In summary, we have shown that for a generic integrable system there exist 
    two different time scales, Ehrenfest time $\tau_\hbar\propto {\hbar}^{-1/2}$ and 
    and quantum revival time  $T_r\propto {\hbar}^{-1}$.  When they are plotted in Fig. \ref{Ed}, they 
    mark up three different regions in the space spanned by $\hbar$ and dynamical evolution time $t$. 
    In the classical region, a narrow quantum wave packet does not spread much and its center follows the classical particle
    trajectory. In the classical ensemble region, a quantum wave packet can be regarded as a classical ensemble distribution 
    in phase space. In the quantum region,  quantum revival occurs and the quantum dynamics can not even be approximated 
    with classical ensemble. 
    
    We call Fig. \ref{Ed} Ehrenfest diagram for two reasons. The first is to honor Ehrenfest for his pioneering work 
    on quantum-classical correspondence~\cite{ehrenfest1927bemerkung}. The second and more important reason is
    that we expect the prominent feature, three different regions marked up by two different time scales, in Fig. \ref{Ed} 
    to be generic. Even for chaotic systems, this feature is expected to persist; the difference is that the Ehrenfest time
    becomes logarithmic  and the quantum revival time will be replaced by other quantum times that scale with $\hbar$ 
    differently. For example, for quantum kicked rotor, the second time is the time scale for dynamical localization
    or quantum resonance and it scales as $\hbar^{-2}$~\cite{Izrailev90}.  We may call this second time scale quantum time. 
    We note that this quantum time in our integrable systems is not Heisenberg time: 
    as indicated in Eq.(\ref{phase}), the quantum revival 
    comes from the second-order terms in the eigen-energy expansion. 
    
    It would be very interesting to see how this kind of Ehrenfest diagram evolves when a system changes from integrable
    to chaotic. It is not clear how Ehrenfest time changes from square root to logarithmic. For a chaotic system, 
    the quantum revival time is likely exponentially long, so the quantum time in the chaotic system must have a 
    different cause. It is not clear what the cause is or whether this cause may change from system to system.  
   
    \acknowledgements
We acknowledge helpful discussion with Yuan Fang on plotting Figs.\ref{phase1}\&\ref{phase2}. 
This work was supported by the The National Key Research and Development Program of China (Grants No.~2017YFA0303302) and the National Natural Science Foundation of China (Grants No.~11334001 and No.~11429402). 
        

\end{document}